\documentclass{ws-procs9x6}

\begin{document}

\newcommand{\earth}{\oplus}

\title{The First Lunar Laser Ranging Constraints on Gravity Sector SME Parameters}

\author{J.B.R.\ BATTAT, J.F.\ CHANDLER AND C.W.\ STUBBS}

\address{Harvard-Smithsonian Center for Astrophysics \\
60 Garden Street, \\
Cambridge, MA 02138, USA \\
E-mail: jbattat@cfa.harvard.edu}

\maketitle

\abstracts{ 
We present the first constraints on pure-gravity sector Standard-Model
Extension (SME) parameters using Lunar Laser Ranging (LLR).  LLR
measures the round trip travel time of light between the Earth and the
Moon.  With 34+ years of LLR data, we have constrained six independent
linear combinations of SME parameters at the level of $10^{-6}$ to
$10^{-11}$.  There is no evidence for Lorentz violation in the LLR
dataset.
}

\section{Introduction and Motivation}
Two of us (C.W.S. and J.B.R.B.) are members of the Apache Point
Observatory Lunar Laser-ranging Operation (APOLLO), a next-generation
LLR facility, capable of millimeter-precision lunar range measurements
(see the article by T.W. Murphy in these proceedings).  The APOLLO
project was motivated by the realization that an order-of-magnitude
improvement in fundamental physics constraints (e.g. equivalence
principle, gravitomagnetism, gravitational $1/r^2$ law and $\dot{G}$,
to name a few) could be achieved with straightforward improvements to
the standard LLR apparatus.

With the recent description of the pure-gravity sector of the
SME,\cite{BK2006} we learned that LLR can also provide incisive
constraints on Lorentz Violation.  The predicted LLR observable under
Lorentz Violation is a periodic perturbation to the Earth-Moon range
with the leading order effects occuring at four distinct frequencies:
$2\omega$, $\omega$, $2\omega-\omega_0$ and $\Omega_{\earth}$.  Here
$\omega$ is the lunar orbital (sidereal) frequency, $\omega_0$ is the
anomalistic lunar orbital frequency and $\Omega_{\earth}$ is the mean
Earth orbital (sidereal) frequency.  Although the APOLLO program is
still in the data collection phase, there are more than three decades
of freely-available archival LLR data on a public archive.\cite{ILRS}
In this article, I present the SME parameter constraints that result
from our analysis of archival LLR data.  These are the first LLR-based
constraints on pure-gravity SME parameters.

\section{The LLR Dataset and Analysis Software}
LLR measures the time of flight of photons between a telescope on the
Earth and corner cubes on the lunar surface.\cite{WTB2005} LLR data is
typically presented in ``normal points'' which are typically generated
from a few to a hundred lunar signal photons collected over a span of
1 to 5 minutes.  Our analysis makes use of archival data from
September 1969 through December 2003.

In the analysis of LLR normal points, a set of residual ranges is
computed by subtracting the model's predicted range from the observed
range.  The range sensitivity with respect to each model parameter
(the partial derivatives) is also computed at the time of each normal
point.  The residuals and the partial derivatives are then used to
compute optimal model parameter values via a weighted linear
least-squares fit.  For our analysis, we used the Planetary Ephemeris
Program (PEP),\cite{PEP} which is currently maintained by one of us
(J.F.C.).  To our knowledge, it is the only publicly available LLR
analysis software.

Typically, the lunar range model is formulated in the parametrized
post-Newtonian (PPN) framework,\cite{willNordtvedt} which permits
model-independent constraints on metric theories of gravity.  At
present, no ephemeris models explicitly incorporate SME parameters.
You can, however, think of these models as implicitly including the SME
parameters but with values pegged at zero ({\em i.e.} no SME
perturbation to the lunar orbit).  It is therefore only necessary to
compute, by hand, the partial derivative of range with respect to each
SME model parameter (see Table \ref{tab:smepartials}).  The analysis
code can then provide SME parameter adjustments simultaneously with
the other model parameters.  A covariance matrix including the
correlations between the SME parameters and all other model parameters
is also produced.

The main drawback to this approach is that one cannot perform an
iterative analysis in which one takes the best-fit model parameter
values and uses them to re-integrate the equations of motion to refine
the model parameter values.  We accept this limitation because the
non-SME model parameters have been highly refined through iterative
solutions over the past several decades and so the solution sits
firmly in the linear regime already.  Furthermore, the addition of the
SME parameters preserves the linearity because the lunar range is
strictly linear in the SME parameters (see Table 2 of
Ref. \refcite{BK2006}), so no iteration is necessary.

\begin{table}[t]
\tbl{SME parameter partial derivatives.  Symbols used here are explained in Ref. 1.}
{\footnotesize
\begin{tabular}{cl}
\hline
SME Parameter & Partial Derivative of Lunar Range with Respect to SME Parameter\\
\hline
$\bar{s}^{11}-\bar{s}^{22}$ & 
  $-\frac{r_{0}}{12}\cos\left(2\omega t+2\theta\right) - 
   \frac{\omega er_{0}}{16\left(\omega-\omega_{0}\right)}\cos\left[\left(2\omega-\omega_{0}\right)t+2\theta\right]$ \\
$\bar{s}^{12}$ & 
  $-\frac{r_{0}}{6}\sin\left(2\omega t+2\theta\right) -
   \frac{\omega er_{0}}{8\left(\omega-\omega_{0}\right)}\sin\left[\left(2\omega-\omega_{0}\right)t+2\theta\right]$ \\
$\bar{s}^{02}$ & 
  $-\frac{\omega\left(\delta m\right)v_{0}r_{0}}{M\left(\omega-\omega_{0}\right)}\cos\left(\omega t+\theta\right)$\\
$\bar{s}^{01}$ & 
   $\frac{\omega\left(\delta m\right)v_{0}r_{0}}{M\left(\omega-\omega_{0}\right)}\sin\left(\omega t+\theta\right)$\\
$\bar{s}_{\Omega_{\earth},c}$ & 
   $V_{\earth}r_{0}\left(\frac{b_{1}}{b_{2}}\right)\cos\left(\Omega_{\earth}t\right)$ \\
$\bar{s}_{\Omega_{\earth},s}$ & 
   $V_{\earth}r_{0}\left(\frac{b_{1}}{b_{2}}\right)\sin\left(\Omega_{\earth}t\right)$ \\
\hline
\end{tabular}\label{tab:smepartials} }
\vspace*{-13pt}
\end{table}

\section{Systematic Errors}
The solar system is complex.  Predictions of the lunar range rely on
models of planetary and asteroid positions, gravitational harmonics of
the Sun, Earth and Moon and various relativistic and non-gravitational
effects (to name a few).  Solar system models have many hundreds of
parameters that influence the Earth-Moon range time.  There are strong
correlations between model parameters.  As a result, solutions will
suffer from systematic errors in model parameter estimates that can
dominate the formal errors reported by the least-squares analysis. In
this work, we account for the underestimation of model parameter
uncertainties by scaling the formal parameter errors reported by the
least-squares analysis by a uniform factor, $F$.  This is numerically
equivalent to uniformly scaling the uncertainty of each normal point
by $F$.  Essentially, we uniformly down-weight the data.

The $F$ factor is empirically determined by holding the SME parameter
values at zero but allowing the PPN values $\beta$ and $\gamma$ to
vary (see Ref. \refcite{willNordtvedt} for an explanation of $\beta$
and $\gamma$).  We know from existing
experiments\cite{betaGammaConstraints} that these parameters are
consistent with their General Relativity values ($\beta=\gamma=1$) to
within a part in $10^3$ or better.  We find that we require $F=20$ to
ensure that we are in accord with these earlier results.

\section{SME Parameter Constraints and Verification}
\begin{figure}[t]
  \centerline{\epsfxsize=3.1in\epsfbox{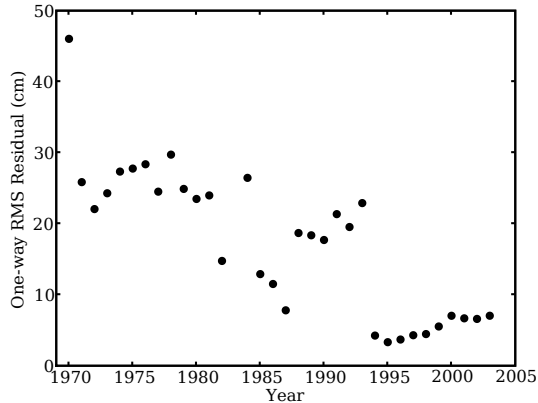}}   
\caption{The agreement between model and data for the first 34 years of 
LLR.
 \label{fig:omcrms}}
\end{figure}

We constrain the SME parameters under the assumption that General
Relativity is not violated (we set $\beta=\gamma=1$).  The resulting
parameter estimates and their realistic errors (the formal errors
scaled by $F=20$) are reported in Table \ref{tab:smeconstraints}.  All
SME parameters are within $1.5F\sigma$ of zero.  There is no evidence
for Lorentz violation in the LLR data.  The fit quality is shown in
Fig. \ref{fig:omcrms}.

To verify our implementation of the partial derivatives of lunar range
with respect to the SME parameters, we generated, by hand, a perturbed
LLR normal point data set by setting
$\bar{s}^{11}-\bar{s}^{22}=9\times10^{-10}$, a $10F\sigma$ deviation
from the best-fit value of this parameter.  A fit to this data
recovers the perturbation: $\bar{s}^{11}-\bar{s}^{22}=[(1+9) \pm
0.9]\times10^{-10}$ with the other SME parameters unchanged.

\section{Conclusions and Future Prospects}
We have analyzed 34+ years of LLR data and have derived constraints on
six SME parameters combinations (see Table \ref{tab:smeconstraints}).
We find no deviation from Lorentz Symmetry at the $10^{-6} - 10^{-11}$
level.  This work provides the first LLR-based constraints of SME
parameters.

There are several ways in which these constraints could be improved.
First of all, by incorporating auxiliary solar system data
(e.g. planetary radar ranging) model parameter correlations can be
reduced, and $F$ decreased.  This would allow for tighter constraints
on the SME parameters using the same LLR dataset.  In addition, APOLLO
data, which is about 10 times more precise than the archival data,
will soon be ready for analysis.  With this improved dataset, we will
further tighten the SME parameter constraints.

\begin{table}[t]
\tbl{SME parameter estimates and their realistic (scaled) uncertainties ($F\sigma$) with $F=20$.  }
{\footnotesize
\begin{tabular}{cr}
\hline
{} &{}\\[-1.5ex]
Parameter & Estimate \\[1ex]
\hline
{} &{}\\[-1.5ex]
$\bar{s}^{11}-\bar{s}^{22}$  & $\left( 1.3 \pm 0.9 \right) \times 10^{-10}$  \\[1ex]
$\bar{s}^{12}$               & $\left( 6.9 \pm 4.5 \right) \times 10^{-11}$  \\[1ex]
$\bar{s}^{02}$               & $\left(-5.2 \pm 4.8 \right) \times 10^{-07}$  \\[1ex]
$\bar{s}^{01}$               & $\left(-0.8 \pm 1.1 \right) \times 10^{-06}$  \\[1ex]
$\bar{s}_{\Omega_\earth c}$  & $\left( 0.2 \pm 3.9 \right) \times 10^{-07}$  \\[1ex] 
$\bar{s}_{\Omega_\earth s}$  & $\left(-1.3 \pm 4.1 \right) \times 10^{-07}$  \\[1ex]
\hline
\end{tabular}\label{tab:smeconstraints} }
\vspace*{-13pt}
\end{table}

\section*{Acknowledgments}
This work developed from discussions with our colleagues including
E. Adelberger, Q. Bailey, J. Davis, A. Kosteleck\'{y}, J. Moran,
T. Murphy, I. Shapiro and M. Zaldarriaga.  We would also like to thank
the staff observers at LLR stations, and the ILRS for the
infrastructure that supports the normal point distribution.
J.B.R.B. acknowledges financial support rom the ASEE NDSEGF, the NSF
GRFP and Harvard University.  We also acknowledge the generous
financial support of the National Science Foundation (grant
PHY-0602507).

\end{document}